\documentclass[12pt,preprint2]{aastex}

\usepackage{epsfig}

\shorttitle{Carbonates in space}
\shortauthors{Posch et al.}

\begin{document}

\title{Carbonates in space -- \\
The challenge of low temperature data\footnote{Based on observations 
with ISO, an ESA project with instruments funded 
by ESA Member States (especially the PI countries: France, 
Germany, the Netherlands and the United Kingdom) and with the 
participation of ISAS and NASA.}}
\author{Th. Posch, A. Baier}
\affil{Institut f\"ur Astronomie, T\"urkenschanzstra{\ss}e 17, A-1180 Wien, Austria}
\email{posch@astro.univie.ac.at}

\and

\author{H. Mutschke}
\affil{Astrophysikalisches Institut,
Schillerg\"asschen 2-3, D-07745 Jena, Germany}

\and

\author{Th. Henning}
\affil{Max-Planck-Institut f\"ur Astronomie (MPIA), K\"onigstuhl 17, 
D-69117 Heidelberg, Germany}


\begin{abstract}

Carbonates have repeatedly been discussed as possible carriers of stardust emission bands. 
However, the band assignments proposed so far were mainly based on room 
temperature powder transmission spectra of the respective minerals.
Since very cold calcite grains have been claimed to be present in 
protostars and in Planetary Nebulae such as NGC 6302, the changes 
of their dielectric functions at low temperatures are relevant from 
an astronomical point of view.

We have derived the IR optical constants of calcite and dolomite from 
reflectance spectra -- measured at 300, 200, 100 and 10\,K -- and calculated
small particle spectra for different grain shapes, with the following results:
i) The absorption efficiency factors both of calcite and dolomite are extremely
dependent on the particle shapes. This is due to the high peak values of the
optical constants of CaCO$_3$ and CaMg[CO$_3$]$_2$.
ii) The far infrared properties of calcite and dolomite 
depend also very significantly on the temperature. Below 200\,K,
a pronounced sharpening and increase in the band strengths of the FIR resonances occurs.
iii) In view of the intrinsic strength and sharpening of the $\sim$44\,$\mu$m 
band of calcite at 200--100\,K, the absence of this band -- inferred from 
Infrared Space Observatory data -- in PNe requires dust temperatures below 
45\,K. 
iv) Calcite grains at such low temperatures can account for the
`92'\,$\mu$m band, while our data rule out dolomite as the carrier 
of the 60--65\,$\mu$m band.

The optical constants here presented are publicly available
in the electronic database 
 {\texttt{http://www.astro.uni-jena.de/Laboratory/OCDB}.}

\end{abstract}



\section{Introduction}

In the 1970s, carbonates have been proposed as carriers of mid-
infrared features detected in astronomical spectra, such as the 11.3\,$\mu$m 
emission band seen in NGC 7027 and in the 'Red Rectangle'
(Gillet, Forrest \& Merrill 1973; Bregman \& Rank 1975;
\cite{Breg77}).

In parallel to that, reflectance spectra of calcite (CaCO$_3$) and magnesite 
(MgCO$_3$) have been measured in the laboratory. \cite{H70} did the measurements
in the whole mid- and far-IR range for calcite, magnesite and dolomite 
(CaMg[CO$_3$]$_2$)), while \cite{P76} examined calcite and magnesite 
in the 5--50\,$\mu$m region. All these measurements have been carried 
out at room temperature.

Carbonate grains have been found in interplanetary dust particles (IDPs).
\cite{San86} and \cite{TB86} reported the identification of carbonate minerals
in IDPs both by a chemical dissolution experiment and by transmission electron
microscopy. As shown by \cite{TB86}, the mineralogy of the IDP carbonates is
essentially that of breunnerite -- an Mg-rich solid solution of magnesite 
and siderite (FeCO$_3$). Solid solutions of Ca- and Mg-carbonates are rare 
in the IDPs examined so far (see also \cite{JB01}). It is noteworthy that 
the IDPs show a 6.8\,$\mu$m absorption feature which is probably due 
to the stretching vibration of carbonate anions (Sandford 1986).
It has to be noted however, that there are doubts on this identification 
because the carbonate band at 11.4\,$\mu$m was not detected in a
study on five IDPs (Quirico et al.\ 2000). A band similar to the 6.8\,$\mu$m 
feature is supposedly seen in the spectra of the ``Deep Impact'' ejecta 
from comet 9P/Tempel 1 measured with the {\em Spitzer}\/ IRS spectrometer 
(Lisse et al.\ 2006). In these spectra, also bands at 11-12 and 
13-14\,$\mu$m have been assigned to carbonates.

While the identification of the 11.3\,$\mu$m band with carbonates 
(\cite{Breg77}) has been confuted by McCarthy, Forrest \& Houck (1978)
and subsequently been replaced by an identification 
with hydrocarbons (\cite{Coh86}), there has been continuous and recently 
even increased interest in carbonates as stardust and planetary dust 
components. Carbonate bands at wavelengths larger than 25\,$\mu$m came 
to the fore: 
\cite{Er00} and \cite{Lel00} suggested calcite and siderite (FeCO$_3$) as 
carriers of aerosol absorption bands of the Martian atmosphere {\em inter 
alia}\/ at 26.5, 30--31 and 43.5\,$\mu$m. 

\cite{K02aunda} and \cite{K02nat}, by a careful analysis of the 2.4--200\,$\mu$m
data of the Planetary Nebula NGC 6302 taken with the Infrared Space Observatory
(ISO), were able to show that the previously unidentified `92'\,$\mu$m emission 
band and the 62\,$\mu$m component of the 60\,$\mu$m emission complex in this 
nebula can possibly be attributed to calcite and dolomite, respectively. According to their model fit, the carbonates belong to the cold (30--60\,K) 
dust component of this nebula. 

\cite{Chia05} studied the 60--180\,$\mu$m ISO spectra of more than 30 protostars 
and detected a solid state feature peaking between 90 and 110\,$\mu$m in 17 
of these spectra. Again, calcite has been suggested as the feature carrier, 
even though the accompanying characteristic resonance features of calcite
could not be detected in the respective sources for different reasons
(partly related to the sensitivity of ISO's detectors). For two star forming 
regions -- the Carina Nebula and Sharpless 171 -- this band had already been 
discovered by \cite{OO03} but had been ascribed to onion structured carbon grains.  
The variation in the band position has been explained by Chiavassa et al.\  
by varying amounts of other than calcium ions (e.g.\ Mg or Fe ions) 
in the carbonatic carrier.

The astronomical observations summarized so far indicate a need for more detailed investigations in two respects: (a) measurement of IR properties below 0\,K and (b) investigation of variations in the band profiles due to particle shape and size effects. Both is best achieved on the basis of determining optical constants for low temperatures. Furthermore, this approach makes it possible to get rid of the possible influence of the matrix effect which may have influenced part of the previous comparisons between laboratory and observational data on carbonates. 

Given that it is unclear how carbonates can form in a circumstellar
environment, \cite{FG05} examined the formation of calcium carbonate in an
oxygen-rich stellar wind (C/O $<$ 1) by thermodynamic calculations, assuming the
formation of calcite from Ca, H$_2$O and CO. They came to the conclusion
that calcite -- if at all -- would rather form in Asymptotic Giant Branch
(AGB) stars with {\em low}\/ mass loss rates (approximately 1 $\times$ 10$^{-6}$ M$_{\odot}$/y) than in AGB stars in their superwind phase (of which planetary nebulae like NGC 6302 are the remnants however). But also in low-mass-loss-rate AGB stars, less than 1\% of the total amount of dust would be carbonate dust. According to \cite{FG05}, it is mainly the relatively low condensation temperature of carbonates which makes their condensation in circumstellar environments so inefficient: for this temperature range to be reached in an expanding circumstellar shell, the dust must have a substantially larger 
distance from the central star (and, accordingly, a substantially 
reduced material density) than for the silicates.

{ Toppani et al.\ (2005), however, suggest a possibly (more) efficient
formation mechanism of carbonates in the Post-AGB phase of stellar
evolution. Based on laboratory experiments, they predict carbonate formation
by reactions between a comparatively hot silicate gas and a 300-500\,K 
H$_2$O-CO$_2$-rich gas under conditions far from local thermal equilibrium
(LTE). They assume this non-LTE carbonate formation to take place when high-velocity winds of forming PNe interact with dense remnants of the 
slow-wind-AGB phase.}

A major goal of our study is to contribute to an assessment of the validity 
of the `carbonate hypotheses' from the spectroscopic point of view.

The structure of the present paper is the following. In \S 2, we describe the derivation of optical constants of CaCO$_3$ and CaMg[CO$_3$]$_2$ from reflectance measurements, both for room temperature and for cryogenic temperatures. In \S 3, absorption efficiency spectra of small calcite and dolomite grains are presented. Finally, in \S 4, we compare these small particle 
spectra with infrared bands of astronomical objects which have been attributed 
to carbonates in recent studies. For the case of calcite, we present new 
constraints on the particle shapes and temperatures which are necessary to 
reconcile the `92'\,$\mu$m band assignments by \cite{K02aunda} and \cite{Chia05}
with currently available laboratory data.


\section{Reflectance spectroscopy and de\-rivation of the optical 
constants \label{s:OC}}

\subsection{Crystallographic properties and sample preparation}

Calcite and dolomite both belong to the trigonal crystal class; hence, they
are both uniaxial, and polarized reflectance measurements are required to 
determine their optical constants.\footnote{Interestingly, the phenomenon of 
{\em birefringence}\/ has been first observed in calcite crystals.} 
The space group of calcite is R$\bar{3}$c, while dolomite belongs to the 
space group R$\bar{3}$ and has a reduced symmetry compared to its chemically 
simpler relative. While the calcium ions occupy
crystallographically equivalent sites in CaCO$_3$, half of the places occupied
by Ca ions in calcite are occupied by Mg ions in dolomite. The reduced symmetry 
of dolomite compared to calcite leads to the occurrence of an additional
IR band (which is located close to 33\,$\mu$m) in its spectrum.
For more details on the lattice structure of Mg and Ca carbonates, see
\cite{H70} and references therein (e.g. \cite{W64}).

In order to perform polarized reflectance spectroscopy, planes containing the
respective crystal's c-axes need to be found or to be produced. This
is a difficult task in the case of ordinary CaCO$_3$ or CaMg[CO$_3$]$_2$ 
rhombohedra, since none of their surfaces fulfill this condition. The c-axis
runs diagonally through the obtuse corners of cleaved rhombohedra. 

In the case of our dolomite rhombohedron -- a 2$\times$2$\times$2\,cm$^3$ 
crystal from Eugui, Navarra, Spain --, it was therefore necessary to cut it in order 
to produce a c-parallel surface, which has been polished subsequently.
Polarized reflectance measurements on the polished surfaces showed no `mixing'
of the ordinary and extraordinary ray, which proved the cut to be parallel 
to the c-axis indeed. 
In the case of our calcite crystal -- a 4$\times$2$\times$1\,cm$^3$ 
fragment of a prism found at Sweetwater, Missouri, USA -- no cut was necessary 
because this individual crystal had a relatively rare stature, its long edges 
being grown parallel to the symmetry axis. Hence, polishing one of the lateral
surfaces of this crystal was sufficient to obtain a large measuring spot
containing the c-axis. Indeed, polarized spectroscopy on that spot did not 
show significant traces of mixing of the two principal orientations of the electric
field vector relative to c (E$\|$c, E$\bot$c; see also below, Fig.\ 
\ref{f:Ref_Calc}).


\subsection{Reflectance spectroscopy}

The reflectance spectra of our two samples have been measured
at near-normal incidence using an FTIR spectrometer Bruker 113v in the
wavelength range 2--120\,$\mu$m at a resolution of 2\,cm$^{-1}$. Gold 
mirrors have been used as references. A turnable grid polarizer on 
polyethylene substrate positioned at the entrance of the sample compartment 
was used to select polarization directions parallel and normal to the 
plane of incidence.

\begin{figure*}[htbp]
\epsscale{1.45}
\plotone{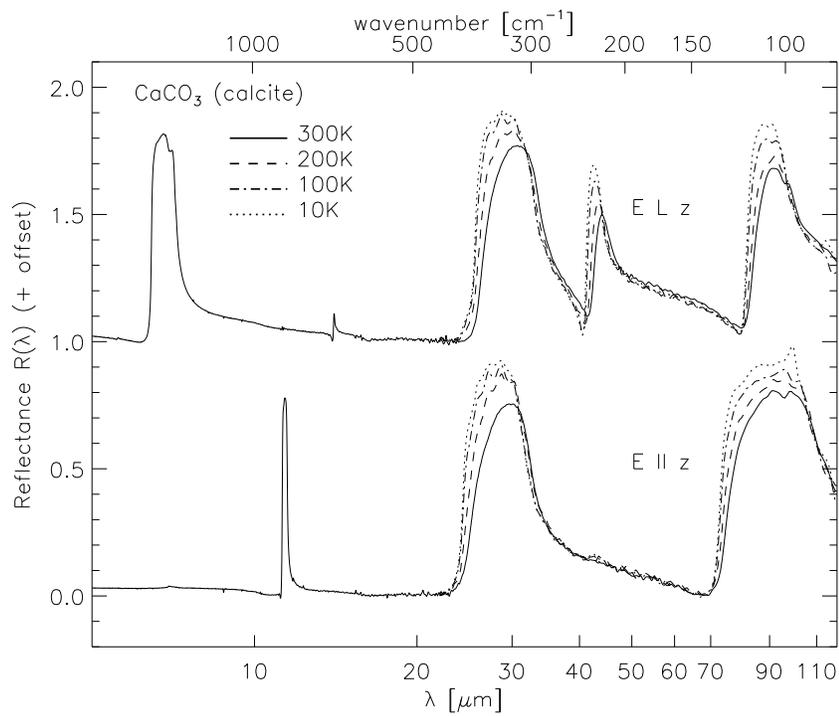}
\caption{Polarized reflectance spectra of calcite for T\,=\,300 K
   (solid line), 200 K (dashed), 100 K (dash-dotted) and
   10 K (dotted line).
   The spectra for E$\bot$c were shifted vertically by an 
   amount of 1.0.
   \label{f:Ref_Calc}}
\end{figure*}

Fig.\ \ref{f:Ref_Calc} shows reflectance spectra of calcite in the 
5--120\,$\mu$m range which comprises all IR bands of this mineral. In the
case of the reflectance for the ordinary ray (E$\bot$c), an offset of 1.0
has been added to all reflectance values. The different linestyles correspond
to the different temperatures (10--300\,K) as indicated.

For the ordinary ray, five reflection bands can be discerned in the
infrared spectrum of calcite, which are centered at 6.8, 14.0, 30.7,
44.0 and 91.4\,$\mu$m for room temperature.
For the extraordinary ray, the respective reflectance band positions
amount to 11.4, 29.6 and 91.4\,$\mu$m (again for room temperature).

\begin{figure*}[htbp]
\epsscale{1.45}
\plotone{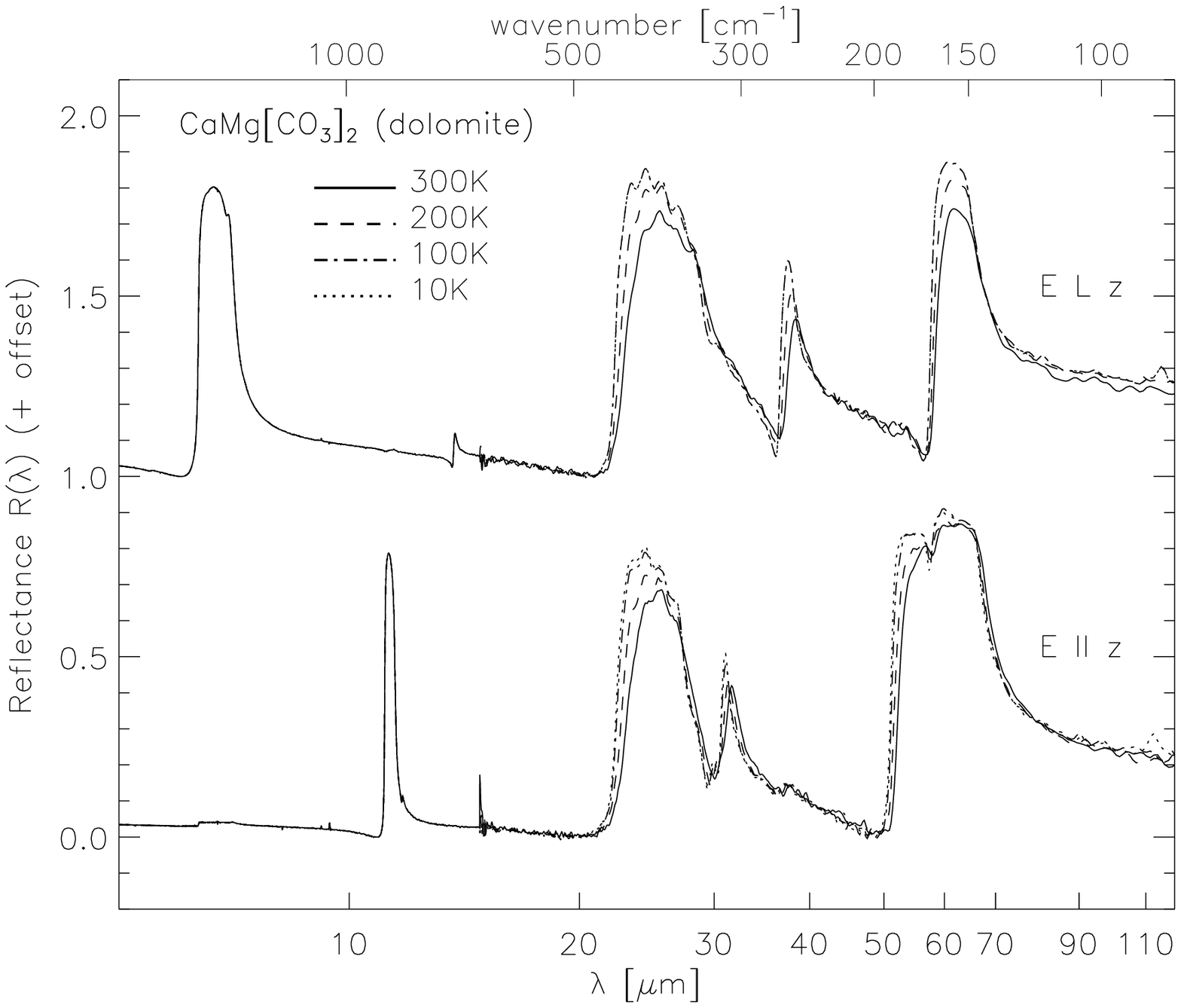}
\caption{Polarized reflectance spectra of dolomite for the same
   series of temperatures as in the case of calcite.
   \label{f:Ref_Dol}}
\end{figure*}

In Fig.\ \ref{f:Ref_Dol}, the reflectance spectra of dolomite are plotted
on the same scale as for calcite. Dolomite has a slightly more complex
band pattern than calcite, which is a consequence of its comparatively 
reduced lattice symmetry mentioned above. The reflectance maxima are located
at 6.6, 13.8, 25.5, 38.4 and 61.7\,$\mu$m for the ordinary ray (at T = 300\,K),
while for the extraordinary ray, the respective positions are 11.3, 25.4, 31.6,
56.7 and 60.0\,$\mu$m. 

For the cryogenic measurements, the experimental setup was the following:
the samples have been placed into a continuous-flow liquid-helium 
cryostat with contact-gas cooling (CryoVac KONTI Spectro B) 
equipped with polyethylene windows. It was inserted into the FTIR 
spectrometer's sample compartment at the same position where the room 
temperature reflection measurements had been performed. The cryostat 
windows did not allow measurements at wavelengths shorter than 15\,$\mu$m. 
The cryostat possesses two sample mounts which can alternatingly be 
moved into the infrared beam, thus allowing subsequent measurements on 
the sample and the reference mirror at each temperature. A third measurement 
with the sample being tilted to suppress the reflection was necessary 
(again at each temperature) in order to determine the reflectivity of 
the cryostat windows (``zero signal''), which had to be subtracted from 
both the sample and the reference spectra. The temperature was controlled 
by means of a CryoVac temperature controller using an ohmic heater and 
{two silicon diode temperature sensors connected to the liquid-helium cooled 
walls of the contact gas chamber and to the sample mount, respectively.}
The temperatures have been stable to about $\pm$0.5~K during the measurements. 

Spectra have been taken at 200\,K, 100\,K, and 10\,K during cooling down the samples. 
Two cooling cycles have been carried out for each sample with the two polarizer 
orientations. The wavelength- and temperature-dependent reflectances have been 
calculated by dividing the ``zero-corrected'' sample and reference spectra at each 
temperature.

It is found that both for dolomite and calcite all bands in the above 
indicated wavelength range undergo a shift to shorter wavelengths, 
a bandwidth decrease and band strength increase with decreasing temperatures. 
These effects are especially significant for calcite's 44\,$\mu$m reflectance 
band and for dolomite's 38\,$\mu$m reflectance band. A more detailed discussion 
is given in \S 3.


\subsection{Lorentz oscillator fits and derived optical constants}

The polarized reflectance spectra of calcite and dolomite can be quite 
accurately reproduced by Lorentz oscillator fits, as has been shown, e.g., 
by \cite{H70}. Using the ansatz

\begin{eqnarray}  \label{Lor}
\epsilon(\nu) = \epsilon_{\infty} + \sum_{j} \frac{\Omega_j^2}{TO_j^2 - \nu^2 - i \gamma_j \nu},
\end{eqnarray}

where $\epsilon_{\infty}$ is the dielectric constant for frequencies large 
compared to all infrared wavelengths, $\Omega_j$ is the strength, TO$_j$ the 
transverse optical {phonon} frequency, $\gamma_j$ the damping constant of the $j$-th 
oscillator, respectively.

\begin{figure*}[htbp]
\plotone{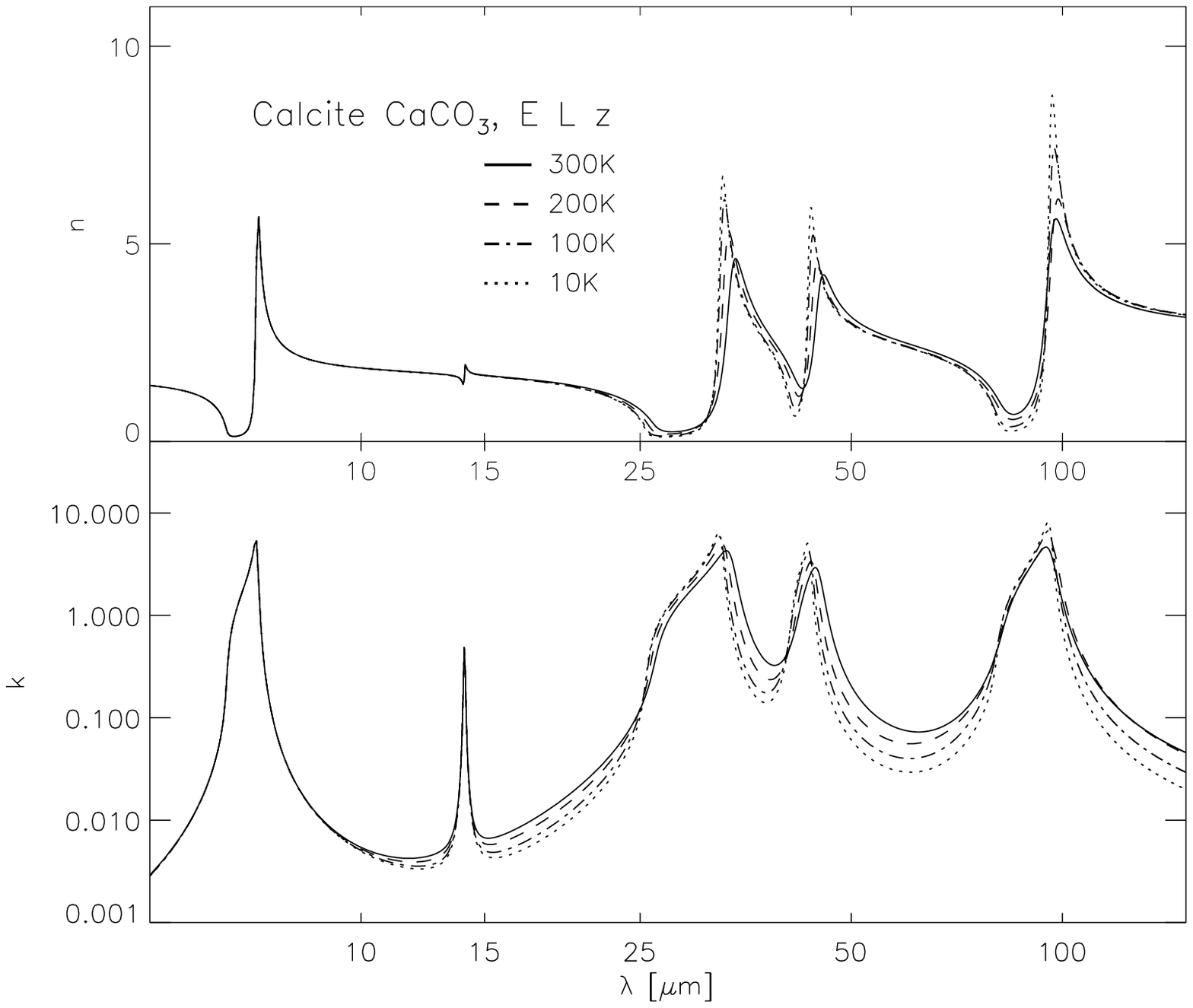}
\plotone{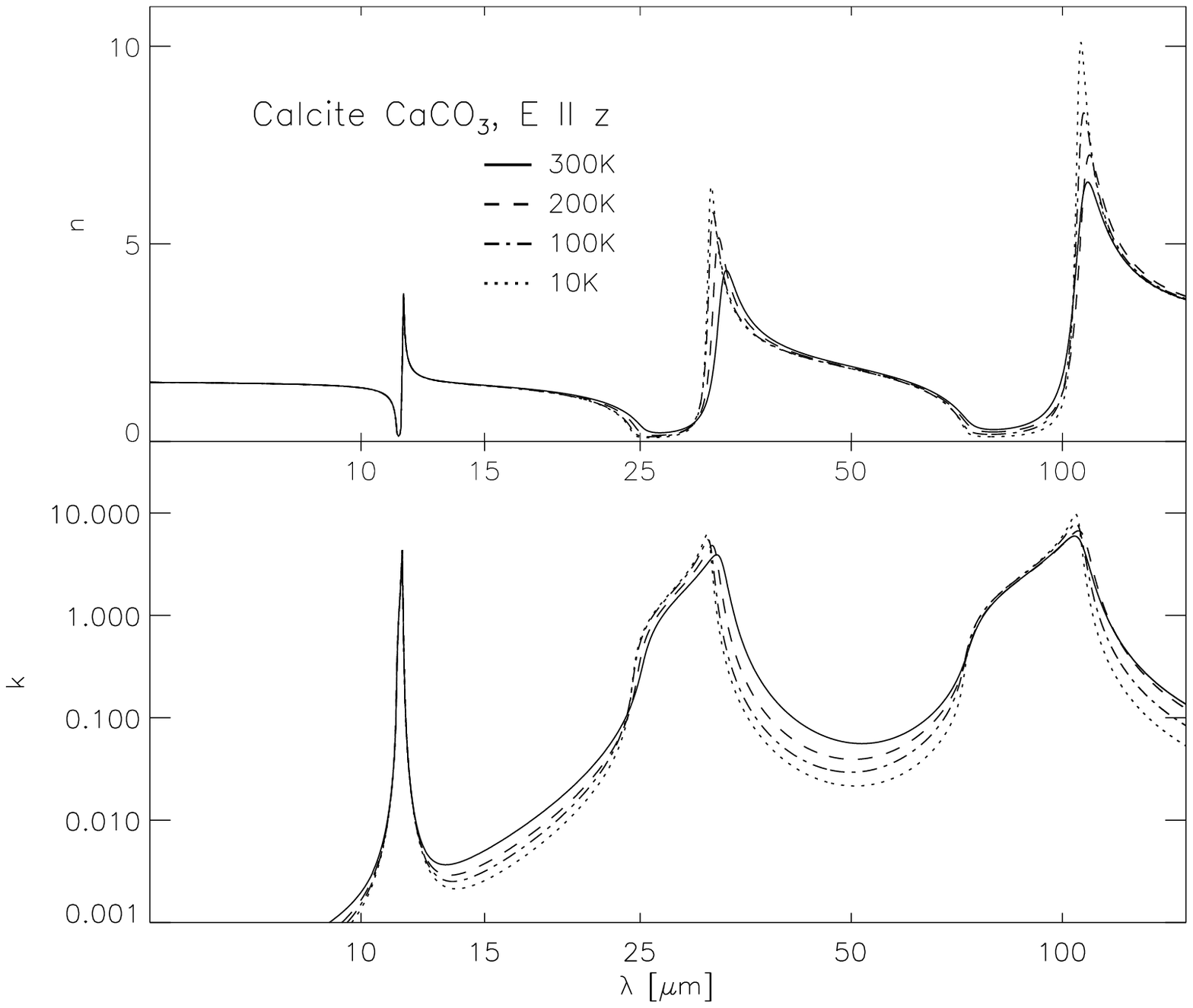}
    \caption{Optical constants of calcite for room temperature, 200\,K, 100\,K
    and 10\,K for the ordinary ray (E$\bot$c) (left) and for the extraordinary ray 
    (E$\|$c) (right). \label{f:nks_Calc}}
\end{figure*}


Figure \ref{f:nks_Calc} shows the optical constants of
calcite which have been derived from the polarized reflectance spectra by
means of Lorentz oscillator fits. The oscillator parameters which were used
for these fits -- five oscillators for the ordinary ray and three for the
extraordinary ray -- are listed in Tab.\ \ref{t:cal-RT-Lor}. 

For dolomite, we only list its oscillator parameters -- six oscillators 
for the ordinary ray and five for the extraordinary ray (see Tab.\ 
\ref{t:dol-RT-Lor}) --, while the optical constants are not shown. 
However, both for calcite and dolomite, the optical constants can be 
retrieved from the electronic database 
{ \texttt{http://www.astro.uni-jena.de/Labo\-ratory/OCDB} },
which also includes graphs.


Apart from the different number of oscillators, another important difference 
between calcite's and dolomite's IR spectra is the absence of any phonon
band for the latter at wavelengths larger than 67\,$\mu$m (or wavenumbers
smaller than 150\,cm$^{-1}$). In other words, it is noteworthy that calcite's
IR bands extend to wavelengths roughly 1.5 times larger than dolomite's
`reddest' band.

For 300\,K, our oscillator parameters are well comparable with those 
derived by \cite{H70} except for some minor bands which are missing 
in their measurements (e.g.\ our `4$_{ER}$' for dolomite),
probably due to a somewhat too small spectral resolution.

\begin{table*}[htbp]
\begin{center}         
         \caption{Lorentz oscillator parameters derived for calcite. 
         `OR' refers to the ordinary ray, `ER' to the extraordinary ray. 
         Where no temperature is indicated as subscript, oscillator 
         parameters have been derived from room temperature measurements.
         For the dielectric background, we chose the 
         value $\epsilon_{\inf, \rm OR}$ = 2.6 in the case of OR and 
         $\epsilon_{\inf, \rm ER}$ = 2.3 in the case of ER. All values 
         are given in units of cm$^{-1}$. 
         \label{t:cal-RT-Lor}}
\begin{tabular}{lccc}
\hline\hline
$j$      & TO$_j$ & $\Omega_j$ &  $\gamma_j$ \\ 
\tableline                                 
1$_{\rm OR}$       & 1407.4 & 1039.6     & 14.8  \\
2$_{\rm OR}$       & 712.0  &  71.4      & 4.0   \\
3$_{\rm OR, 300K}$ & 297.0  & 377.7      & 15.7  \\
3$_{\rm OR, 200K}$ & 302.5  & 385.4      & 12.0  \\
3$_{\rm OR, 100K}$ & 306.0  & 387.0      & 8.87  \\
3$_{\rm OR, 10K}$  & 307.1  & 387.6      & 7.34  \\
4$_{\rm OR, 300K}$ & 222.8  & 206.1      & 11.0  \\
4$_{\rm OR, 200K}$ & 226.5  & 211.2      & 8.35  \\
4$_{\rm OR, 100K}$ & 228.8  & 219.1      & 6.14  \\
4$_{\rm OR, 10K}$  & 229.7  & 219.7      & 4.54  \\
5$_{\rm OR, 300K}$ & 104.7  & 148.1      & 6.00  \\
5$_{\rm OR, 200K}$ & 103.2  & 165.7      & 5.35  \\
5$_{\rm OR, 100K}$ & 103.8  & 158.0      & 3.57  \\
5$_{\rm OR, 10K}$  & 104.2  & 167.6      & 2.45  \\
\tableline
1$_{ER}$      & 872.0  & 250.8      & 2.09    \\
2$_{ER,300K}$ & 306.7  & 368.0      & 16.9    \\
2$_{ER,200K}$ & 312.6  & 372.2      & 11.6    \\
2$_{ER,100K}$ & 317.6  & 377.6      & 9.19    \\
2$_{ER,10K}$  & 318.9  & 378.7      & 7.30    \\
3$_{ER,300K}$ & 94.3   & 200.6      & 7.10    \\
3$_{ER,200K}$ & 93.3   & 203.2      & 5.91    \\
3$_{ER,100K}$ & 94.6   & 203.6      & 4.37    \\
3$_{ER,10K}$  & 95.0   & 202.4      & 2.86    \\
\tableline
\end{tabular}
\end{center}
\end{table*}

\begin{table*}[htbp]
\begin{center}         
         \caption{The set of Lorentz oscillator parameters derived
         for dolomite by fitting the room temperature reflectance
         spectra. All designations have the same meaning as in the
         previous table, but here, we chose 
         $\epsilon_{\inf, \rm OR}$ = 2.7 and 
         $\epsilon_{\inf, \rm ER}$ = 2.4.
          \label{t:dol-RT-Lor}}
\begin{tabular}{lccc}
\hline\hline
$j$      & TO$_j$ & $\Omega_j$ &  $\gamma_j$      \\ 
\tableline                                 
1$_{\rm OR}$       & 1453.8 & 399.2      & 23.8   \\
2$_{\rm OR}$       & 1425.5 & 980.0      & 15.0   \\
3$_{\rm OR}$       & 727.3  & 98.8       & 6.14   \\
4$_{\rm OR, 300K}$ & 348.7  & 459.0      & 22.1   \\
4$_{\rm OR, 200K}$ & 351.5  & 462.5      & 16.0   \\
4$_{\rm OR, 100K}$ & 356.0  & 464.4      & 12.0   \\
4$_{\rm OR, 10K}$  & 355.9  & 464.5      & 11.9   \\
5$_{\rm OR, 300K}$ & 259.1  & 190.6      & 9.86   \\
5$_{\rm OR, 200K}$ & 261.1  & 201.9      & 7.97   \\
5$_{\rm OR, 100K}$ & 263.4  & 201.5      & 5.58   \\
5$_{\rm OR, 10K}$  & 263.4  & 201.4      & 5.54   \\
6$_{\rm OR, 300K}$ & 152.9  & 209.1      & 5.18   \\
6$_{\rm OR, 200K}$ & 152.9  & 214.2      & 3.42   \\
6$_{\rm OR, 100K}$ & 153.5  & 214.9      & 2.26   \\
6$_{\rm OR, 10K}$  & 153.5  & 214.7      & 2.20   \\
\tableline
1$_{ER}$      & 874.9  & 301.0      & 2.98    \\
2$_{ER,300K}$ & 364.8  & 335.8      & 20.4    \\
2$_{ER,200K}$ & 368.5  & 341.8      & 16.3    \\
2$_{ER,100K}$ & 370.7  & 349.8      & 12.7    \\
2$_{ER,10K}$  & 371.5  & 356.4      & 11.9    \\
3$_{ER,300K}$ & 316.8  & 179.5      & 9.13    \\
3$_{ER,200K}$ & 319.4  & 175.6      & 8.24    \\
3$_{ER,100K}$ & 321.3  & 168.1      & 6.05    \\
3$_{ER,10K}$  & 321.6  & 178.0      & 5.05    \\
4$_{ER,300K}$ & 171.9  &  30.5      & 8.76    \\
4$_{ER,200K}$ & 174.0  &  29.2      & 6.42    \\
4$_{ER,100K}$ & 174.4  &  31.1      & 5.56    \\
4$_{ER,10K}$  & 174.8  &  34.4      & 4.97    \\
5$_{ER,300K}$ & 150.3  & 243.2      & 3.35    \\
5$_{ER,200K}$ & 151.2  & 247.5      & 3.29    \\
5$_{ER,100K}$ & 151.4  & 249.8      & 2.31    \\
5$_{ER,10K}$  & 152.2  & 260.4      & 2.17    \\
\tableline
\end{tabular}
\end{center}
\end{table*}


\section{Small particle spectra for different shape distributions \label{s:mie}}

From the optical constants shown and discussed in the previous section, the
absorption cross sections per particle volume -- C$_{abs}$/V -- have been 
calculated for {grains of different shapes. Since the focus of this paper is 
on bands at large wavelengths, we choose the limit of particles being small 
compared to the wavelength (Rayleigh limit, size parameter 
x=2$\pi$a/$\lambda\ll$ 1 and $\vert\epsilon^{1/2}\vert$ x $\ll$ 1,
$a$ being the grain radius or largest semidiameter).

The following grain shapes have been taken into account: spheres (according to Mie 
theory) and continuous distributions of ellipsoids (CDEs). As for the latter, two
different modifications of it have been used: a CDE with equal probability for all
ellipsoid axis ratios (`mean CDE', after \cite{BH83}) and a CDE with maximum 
probability for spherical grains (`weighted CDE', after \cite{Osk92}).

\subsection{Temperature effects for spherical grains}

{Figures \ref{f:Cabs_Cal} and \ref{f:Cabs_Dol} show the volume normalized absorption cross 
sections C$_{abs}$/V calculated for spherical calcite and dolomite grains, respectively, 
depending on temperature. 
The figures demonstrate that the effect of the cooling on the band positions
and band widths is present also in the small particle spectra. Similarly to the 
reflectance spectra, this effect is especially pronounced for the case of the 
$\sim$40\,$\mu$m band of calcite, which shifts from 42.6\,$\mu$m for 300\,K to 
41.4\,$\mu$m for 10\,K, while at the same time, the C$_{abs}$/V peak value increases 
by a factor of 2.4. For the calcite bands around 30 and 80\,$\mu$m, the 
temperature dependence is not equally strong, but still significant.

In the case of CaMg[CO$_3$]$_2$, the tempera\-ture-dependent evolution of the spectra 
is similar apart from the effect that there is only an insignificant change between 
100\,K and 10\,K (cf. Fig.\ \ref{f:Cabs_Dol}). For the additional band in the dolomite 
spectrum (compared to calcite) around 31\,$\mu$m, the temperature dependence is rather 
moderate compared to the other bands. In the wavelength range 
$\lambda$\,$<$\,15\,$\mu$m, the dolomite and the calcite spectra resemble
each other very closely.

\begin{figure*}[htbp]
\epsscale{1.6}
\plotone{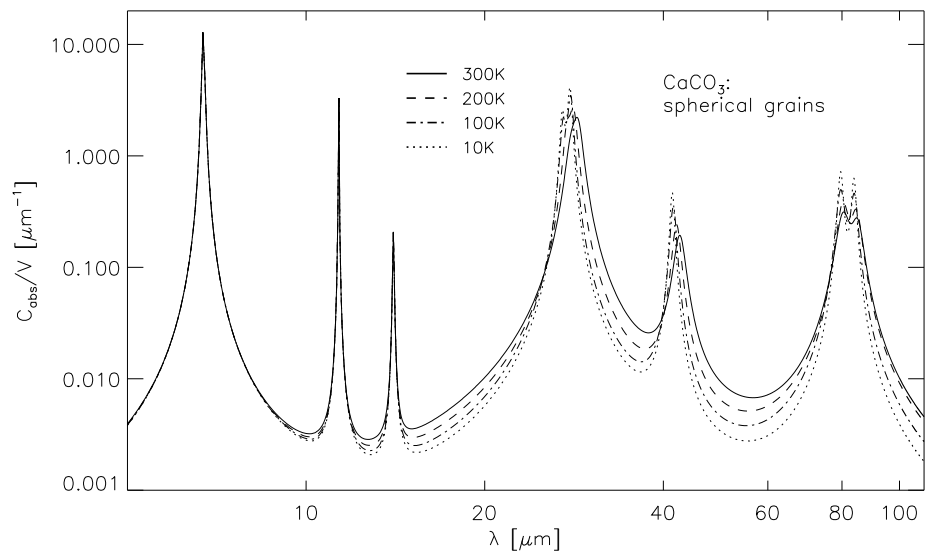}
\caption{Volume normalized absorption cross sections of spherical calcite 
grains for room temperature, 200\,K, 100\,K and 10\,K in the wavelength 
range 5--110\,$\mu$m.
   \label{f:Cabs_Cal}}
\end{figure*}

\begin{figure*}[htbp]
\epsscale{1.6}
\plotone{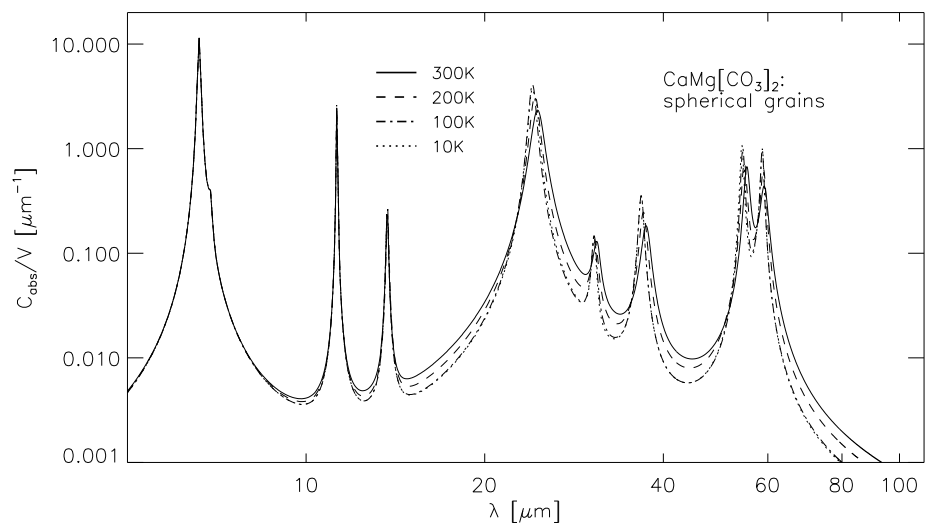}
\caption{Volume normalized absorption cross sections of spherical dolomite 
grains for room temperature, 200\,K, 100\,K and 10\,K.
   \label{f:Cabs_Dol}}
\end{figure*}

\subsection{Grain shape effects at low temperatures}

Due to the high oscillator strengths characterizing most of the infrared bands of both 
calcite and dolomite, the grain shape dependence of the band profiles is very strong. 
For the FIR bands of calcite, this is evident from the comparison of 
Fig.\ \ref{f:Cabs_Cal} with Fig.\ \ref{f:Cabs_Cal_CDE} as well as from the 
peak wavelengths given in Tab.\ \ref{t:shapes}. Calcite's 30\,$\mu$m band 
is shifted to the `red' by about 2\,$\mu$m for a mean CDE compared to the case of 
spherical grains, and at the same time its width is increased by factor of four. 
The 80-90\,$\mu$m double-band of calcite degenerates into a broad single bump 
centered at 90.1\,$\mu$m with two wings for the mean CDE. 
For the weighted CDE case, the maximum of the absorption
efficiency is located at 86.3\,$\mu$m for a grain temperature of 300\,K. 
For the 40\,$\mu$m band, the broadening due to the shape distribution 
is not equally strong as for the 80-90\,$\mu$m band, and the shift of
the band position to longer wavelengths is limited to 1\,$\mu$m for
a given temperature (compared to more than 5\,$\mu$m for the 80-90\,$\mu$m
band). 

\begin{table*}[htbp]
\begin{center}         
         \caption{Positions of the maxima of C$_{abs}$/V for
         calcite grains at different temperatures and for different
         shape distributions
          \label{t:shapes}}
\begin{tabular}{lccc}
\hline\hline
    & 20\,$\mu$m complex    & 40\,$\mu$m & 80--90\,$\mu$m        \\ 
Peak position & [$\mu$m] & [$\mu$m]   & [$\mu$m]                 \\
\tableline 
\textbf{sphere}     &            &            &                  \\
    10\,K     &    27.05, 27.84  &    41.40   &   79.62, 83.77   \\
   300\,K     &    28.58         &    42.64   &   80.56, 84.71   \\
\tableline 
\textbf{weighted CDE}  &                &            &           \\
 10\,K        &         28.18           &    41.75   &   84.91   \\
 300\,K       &         29.13           &    43.09   &   86.25   \\
\tableline 
\textbf{mean CDE} &                &             &               \\
 10\,K        &         30.02      &     42.54   &   90.31       \\
 300\,K       &         30.66      &     43.68   &   90.14       \\
\tableline
\end{tabular}
\end{center}
\end{table*}

Comparing the band properties for different temperatures within a given grain model, 
we find that the shift of the peak position is very prominent in the case of 
the 40\,$\mu$m feature, even for the grain shape distributions. It clearly dominates 
over shape effects for this band, whereas this is not true  
for the 30 and 80--90\,$\mu$m features. In the case of the latter both the shape effects 
are stronger due to the higher oscillator strengths and the temperature effects are 
intrinsically smaller as has been discussed in the previous section. 

\begin{figure}[htbp]
\epsscale{1.0}
\plotone{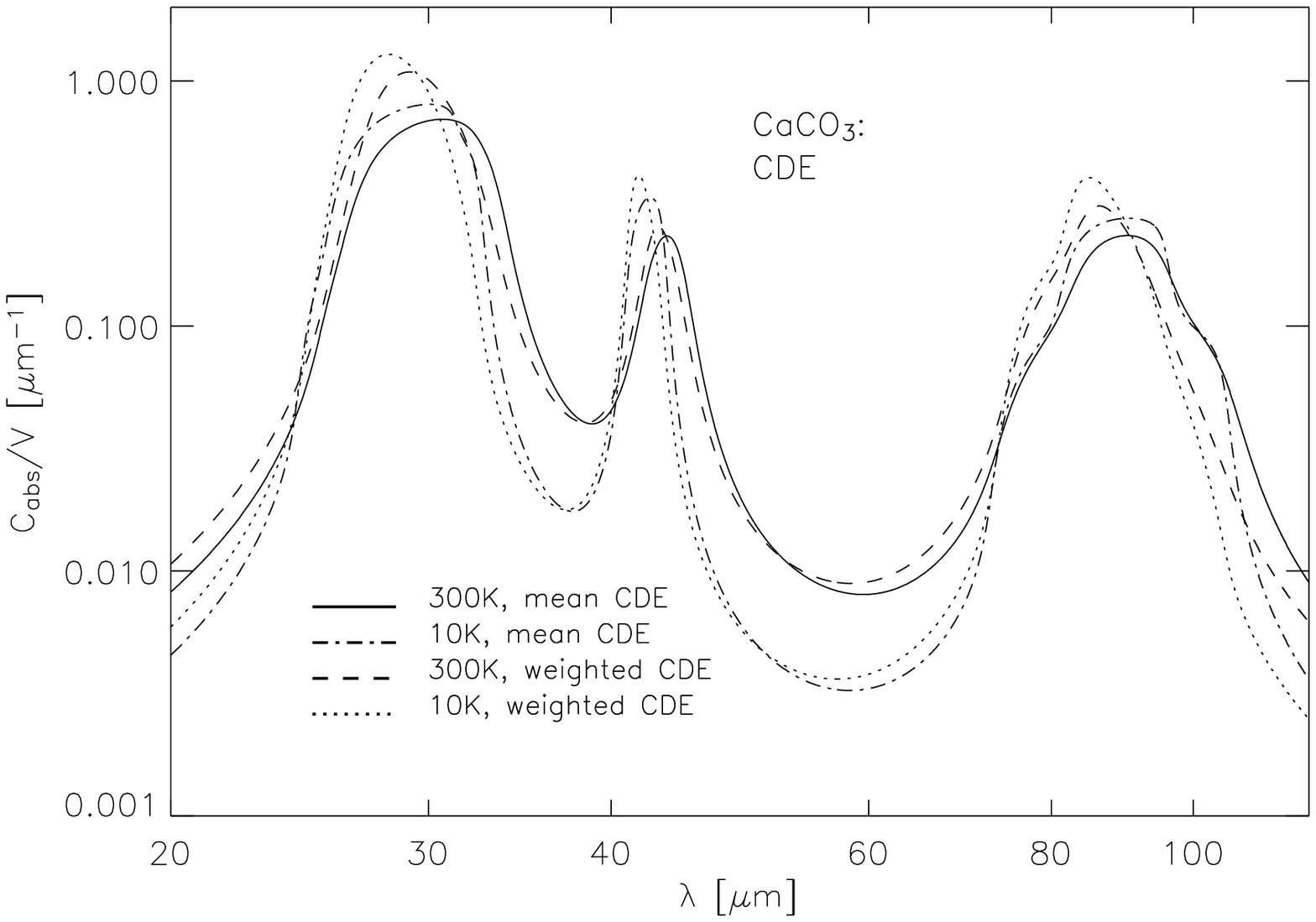}
\caption{C$_{abs}$/V profiles for a distribution of ellipsoidally 
	shaped calcite grains (see text for details). Only the wavelength range
	20--120\,$\mu$m and only the two temperature `extremes'
	T=10\,K and T=300\,K have been depicted here.
    \label{f:Cabs_Cal_CDE}}
\end{figure}

\subsection{Comparison to PE powder spectra}

Since powder spectra of calcite and dolomite -- measured at room 
temperature -- have been utilized by both Kemper et al.\ (2002a) and 
Chiavassa et al.\ (2005) in order to argue for the presence of carbonate dust in 
astronomical objects, it is important to discuss the relation between such 
powder spectra and the small particle spectra based on our optical constants. 
By `powder spectra' or `PE spectra', we understand the transmission spectra 
of sub-micron sized ground powder embedded in polyethylene (PE), 
transformed into absorption cross sections. 

Powder spectra generally depend on various circumstances such as particle size 
and shape, orientation, agglomeration degree, and the refractive index of the 
surrounding medium (e.g.\ \cite{BH83}). 

Fig.\ \ref{f:PE_CDE} shows both the powder spectrum published by Kemper
et al.\ (2002a) -- measured on a sample of unsedimented calcite grains 
(dash-dotted-line) -- and of a sample of calcite grains which had previously 
been subjected to sedimentation in a liquid (acetone), in order to remove large grains (solid line).
The peak of the latter spectrum is shifted by about 2\,$\mu$m to shorter wavelengths with respect to the data published by Kemper et al.\ (2002a), although both have been measured at room temperature. 
{ This is due to the presence of large grains (several $\mu$m in diameter) 
in the unsedimented powder.}
The spectrum of the sedimented powder is in good agreement with the result of the calculation for a {\em mean}\/ CDE with PE environment (see Fig.\ \ref{f:PE_CDE}), indicating the presence of a considerable fraction of grains with 
shapes far away from spherical symmetry.  

Conversely, we find that the mean CDE is a good representation of the shape 
distribution of a real calcite powder with grain sizes within the Rayleigh 
limit. This justifies the use of this shape distribution for the comparison with 
observed data at large wavelengths. In the following (see Sect.\,4.1) we will 
apply this model to simulate dust spectra at low temperatures. 

\begin{figure}[htbp]
\plotone{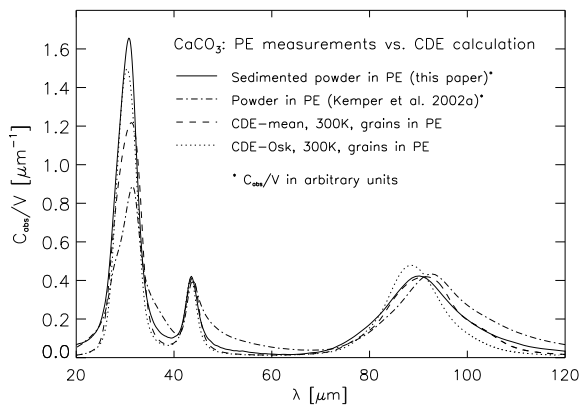}
\caption{Comparison of calcite's FIR bands measured by PE powder transmission
spectroscopy (solid and dashed-dotted lines; see text for details) with the absorption efficiency spectrum calculated from the optical constants for a mean continuous distribution of ellipsoids (dashed) and for a weighted CDE (dotted), 
both in PE.
   \label{f:PE_CDE}}
\end{figure}


\section{Comparison with astronomical observations \label{s:comp}}

The laboratory spectra presented so far are especially relevant for a comparison
with FIR spectra of cold dust in objects such as the planetary nebula NGC 6302. 
Hence, we shall first give a brief review of some key properties of the dust 
spectrum of this object.

Barlow (1997) observed features 65 and 69\,$\mu$m and another possible broad 
band feature in the range of 88-98\,$\mu$m. While the 65\,$\mu$m feature was 
attributed to crystalline ice (see also Waters et al.\ 1996), the 69\,$\mu$m
feature was suspected to be due to crystalline forsterite and the question
for the carrier of the 88-98\,$\mu$m band remained open.

Molster et al.\ (2001) identified the $\sim$65\,$\mu$m band as a blend of diopside 
and crystalline water ice with enstatite. They also confirmed the reality of the 
broad feature around 90\,$\mu$m and suspected that there should be a very cold dust 
component present in the nebula.

Kemper et al.\ (2002a,b) first assigned { the $\sim$65\,$\mu$m feature
to cold dolomite (with additional contributions by diopside and water ice)}
and the broad emission band around 90\,$\mu$m to cold calcite dust. As already 
mentioned above, the temperature of the carbonate dust was assumed to be in the 
30-60\,K range. A mass fraction of less than 0.3\% (for calcite as well as for dolomite) has been derived by these authors.

We retrieved an ISO-SWS and an ISO-LWS spectrum of NGC 6302 from the ISO archive and reduced it by means of the OLP version 10.0.

From the composite spectrum (ranging from 2.4--200\,$\mu$m), we subtracted a combination of Planck functions for temperatures of 30, 55 and 95\,K. The remaining `residual dust emission' is shown in Fig.\ \ref{f:resid1} for the 
35--105\,$\mu$m range; the individual broad peaks of the gray line result from various ice and dust species such as H$_2$O, forsterite, enstatite, diopside, and -- possibly -- carbonates (see Kemper et al.\ 2002a, Tab.\ 2).

\begin{figure}[htbp]
\plotone{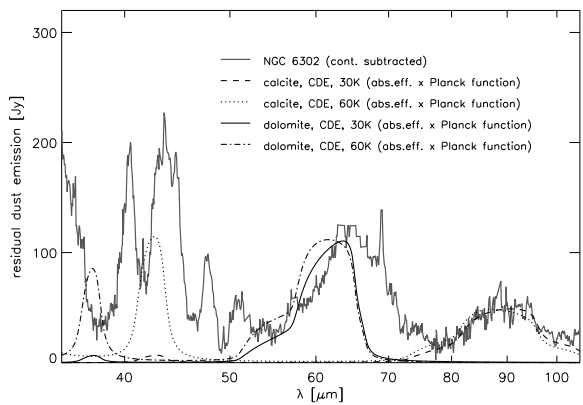}
\caption{Comparison between the residual dust emission of NGC 6302
and synthetic dust emission spectra (for 30\,K and 60\,K) of
calcite and dolomite in the 35-105\,$\mu$m range.
   \label{f:resid1}}
\end{figure}

Our aim has been to examine whether the assignments of the 60-65\,$\mu$m
and `92'\,$\mu$m bands to dolomite and calcite is compatible with the
optical constants and powder spectra presented in Sect.\ 3, and, if so, 
for which temperatures and particle shapes this is the case. For 
answering this questions, the detectability of bands at wavelengths 
between 35 and 45\,$\mu$m depending on the dust temperature also 
plays a key role.

\subsection{The `92'\,$\mu$m band in NGC 6302}

A detailed view of the `92'\,$\mu$m band is given in Fig.\ \ref{f:resid2}.
It can be seen from this figure (when compared to Fig.\ \ref{f:Cabs_Cal_CDE})
that only for a mean CDE particle shape distribution of calcite grains, 
the bandwidth is as large as in the case of the NGC 6302 spectrum.
For a weighted CDE, the bandwidth is smaller by 45\% (at the relevant
cryogenic temperatures). The best fit is achieved by the C$_{abs}$/V function 
for a mean CDE multiplied by a Planck function for a temperature 
of 30\,K. A dust temperature of 60\,K is hardly compatible with the observations, 
both with respect to the `92'\,$\mu$m band profile and with respect to the 
emergence of a $\sim$42\,$\mu$m feature (see dotted lines in Figs.\ 
\ref{f:resid1} and \ref{f:resid2}). The calcite data derived from 
transmission spectroscopy of {\em sedimented}\/ powder (dash-dotted line 
in Fig.\ \ref{f:resid2}) roughly reproduce the observed band profile as 
well. It is noteworthy that the `92'\,$\mu$m band profile of NGC 6302 
obtained by Kemper et al.\ (2002a) is better fitted by the PE spectrum 
of an {\em unsedimented}\/ calcite powder sample which is broader and peaks 
at a larger wavelength. This may be due to the use of different ISO LWS spectra 
(a mean of seven vs.\ a single) or due to some uncertainty in the continuum 
subtraction. 

\begin{figure}[htbp]
\plotone{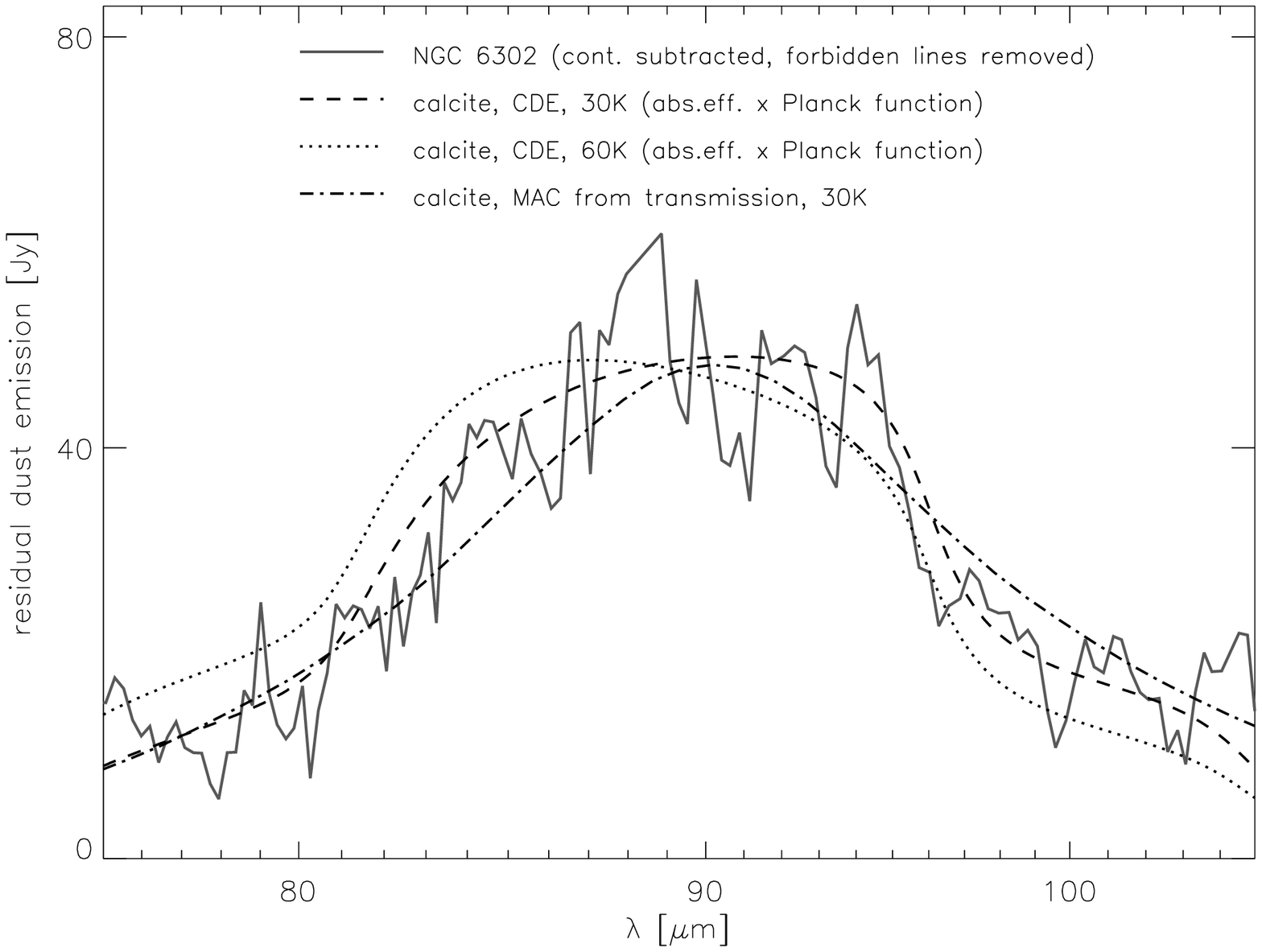}
\caption{Comparison between the resi\-dual dust emission of NGC 6302
and synthetic dust emission spectra of ensembles of ellipsoidally
shaped calcite grains for 30\,K and 60\,K in the 75--105\,$\mu$m range.
   \label{f:resid2}}
\end{figure}

\subsection{The 60-65\,$\mu$m emission complex in NGC 6302}

In Fig.\ \ref{f:resid3} a close-up view of the 60\,$\mu$m emission complex is shown. 
Although the bandwidth of a mean CDE particle shape distribution can reproduce the 
observed bandwidth of the 60\,$\mu$m feature, neither for a mean CDE nor for a 
weighted CDE -- both at a temperature of 45\,K -- the {\em positions}\/ coincide with the 
observed one. The difference between the band positions resulting from
laboratory measurements and the observed band position ($\sim$\,64\,$\mu$m) is
too large as to be compensated by temperature effects -- see also Fig.\
\ref{f:resid1}, where a comparison with dolomite dust at T=30\,K is made. 
Dust temperatures lower than 40\,K would allow to suppress the dolomite 
emission band at $\sim$37\,$\mu$m (see again Fig.\ \ref{f:resid1}), but the 
position of the 60\,$\mu$m feature would still not match. 
Therefore this feature cannot be fit only by dolomite,
while it is not excluded that dolomite contributes to the `blue wing' of
the 60\,$\mu$m emission complex. Our study confirms that the 60-65\,$\mu$m 
band must have other major carriers, e.g.\ diopside as suggested by 
Kemper et al.\ (2002a). A recent publication by Chihara et al.\ 
(2007) shows that a mixture of diopside and {\aa}kermanite (Ca$_2$MgSi$_2$O$_7$) 
at a temperature of 30\,K provides a good match of the 60-65\,$\mu$m feature.

\begin{figure}[htbp]
\plotone{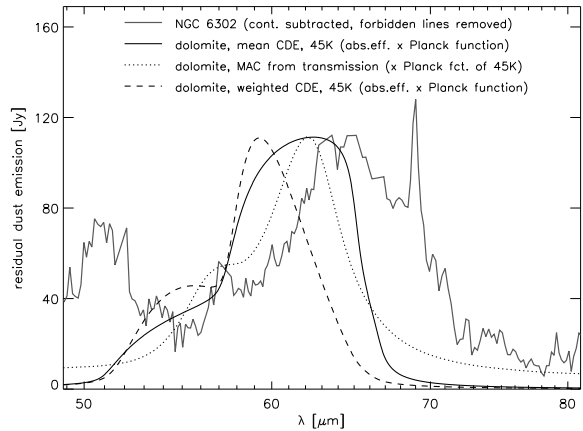}
\caption{Comparison between the residual dust emission of NGC 6302
and synthetic dust emission spectra of ensembles of ellipsoidally
shaped dolomite grains for 30\,K and 50\,K in the 49--81\,$\mu$m range.
   \label{f:resid3}}
\end{figure}

\subsection{The 90--110\,$\mu$m band in protostars}

Chiavassa et al.\ (2005) studied a sample of 32 low and intermediate 
mass protostars (mostly Class 0 and Herbig Ae/Be objects). Out of this 
sample, 17 sources show a broad band feature between 90 and 110\,$\mu$m.
The dust temperature in these objects is found by blackbody fits to the 
continuum to be below 50\,K (except for emission from the innermost regions). 
Calcite is suggested to be a possible carrier of this feature, which seems 
to be supported by the laboratory measurements of Kemper et al.\ (2002a) 
that show a peak at 93\,$\mu$m and a width (FWHM) of 16\,$\mu$m. 

Our data show that for PE powder measurements on small grains and also for 
calculated spectra assuming a mean CDE, the maximum of the `92'$\mu$m 
calcite band is located at 90$\mu$m.
For other particle shape distributions, it is narrower and located 
at much smaller wavelengths, especially for the temperature range relevant 
for PNe and protostars. The transverse optical resonance wavelength 
of mode 5$_{OR,10K}$ of about 96\,$\mu$m (Tab.\ 1) can be considered 
as the maximum peak position for this band (cf.\ \cite{BH83}). The 
feature produced by the 3$_{ER,10K}$ mode with a TO resonance wavelength 
of about 105\,$\mu$m is much broader and therefore comparably weak 
and cannot dominate the calcite grain emission even if alignment 
effects were present.

Hence we conclude that the bands observed for the Herbig Ae/Be objects 
may in principle be compatible with emission from calcite grains, if sufficiently 
extreme shape distributions (much more extreme than the mean CDE) and/or 
large grains are assumed (cf.\ Fig.\ 7 in Chiavassa et al.\ 2005). 
For the Class 0 objects, in the contrary, we confirm that the feature 
characteristics are incompatible with laboratory data for calcite. 
Whether the incorporation of other ions into a carbonate can shift the 
band to still larger wavelengths has to be left open at this point. 


\section{Conclusions \label{s:conc}}

While previous studies on carbonate dust in astronomical environments were
largely based on powder measurements performed at room temperature,
we have on the one hand derived optical constants of calcite and dolomite 
and on the other hand carried out measurements at cryogenic temperatures
(200, 100 and 10\,K). From these optical constants, small particle spectra 
(absorption efficiencies as functions of the wavelength and temperature)
were calculated for different grain shapes.

Since the optical constants of calcite and dolomite reach peak values of 
5--10 (corresponding to maxima of the imaginary part of the dielectric function 
of up to 150), their absorption efficiencies are extremely sensitive to particle
shape effects. For the `92'\,$\mu$m band of calcite, e.g., the negativity 
range of the real part of the dielectric function is as large as 
73--96\,$\mu$m for the ordinary ray (and even 75--105\,$\mu$m for 
the extraordinary ray). The former values approximately indicate
how large the range is in which calcite's `92'\,$\mu$m band can shift due
to particle shape effects, since the ordinary ray is weighted more 
strongly than the extraordinary ray.

At cryogenic temperatures, the FIR bands of calcite and dolomite increase
their intensity, get narrower and shift to shorter wavelengths. The 
temperature-related shift in position is largest for calcite's 40\,$\mu$m
band: in this case, it reaches 0.5\,$\mu$m per 100\,K.

As for the `92'\,$\mu$m feature detected by Kemper et al.\ (2002a) in ISO-LWS
spectra of NGC 6302, this feature can only be assigned to CaCO$_3$ grains
if the assumption of a mean CDE -- with equal probability of all grain shapes 
-- is made.

Furthermore, dust temperatures have to be well below 40\,K in order
to suppress the 30\,$\mu$m and 42\,$\mu$m bands of calcite which accompany
the `92'\,$\mu$m band but are not seen in the spectra of NGC 6302.


\begin{small}
{\acknowledgments

We thank Gabriele Born and Walter Teu\-schel, Jena, for
sample preparation and help with the cryogenic measurements.
Ingrid Hodou\v s kindly provided the ISO-spectrum of
NGC 6302, reduced with the ISO spectral analysis 
package ISAP. HM acknowledges support by DFG grant 
Mu \mbox{1164/5.} TP acknowledges support by the DFG
`Forschergruppe' {\em laboratory astrophysics}\/,
by a grant from the Max-Planck-Society. AB has been supported 
by the Austrian `Fonds zur F\"orderung der wissenschaftlichen
Forschung' (FWF; project number P18939-N16).
Comments by an anonymous referee helped to improve
the original version of this paper.}

\end{small}


\end{document}